\documentstyle[12pt]{article}
\begin{document}
\rightline{NKU-03-SF2}
\bigskip
\begin{center}
{\Large\bf Quasinormal Modes  of  Charged  Dilaton Black Holes 
 in 2 +1 Dimensions 
}

\end{center}
\hspace{0.4cm}
\begin{center}
Sharmanthie Fernando \footnote{fernando@nku.edu}\\
{\small\it Department of Physics \& Geology}\\
{\small\it Northern Kentucky University}\\
{\small\it Highland Heights}\\
{\small\it Kentucky 41099}\\
{\small\it U.S.A.}\\

\end{center}

\begin{center}
{\bf Abstract}
\end{center}

\hspace{0.7cm} 

In this paper,   scalar perturbations of static charged dilaton black holes in 2+1 dimensions are studied. The black hole  in tis paper  is a solution to the low-energy string theory in 2+1 dimensions. Exact values of quasinormal modes for the scalar  perturbations are calculated. For both the charged and uncharged cases, the quasinormal frequencies are pure-imaginary leading to purely damped modes for the perturbations.


\section{Introduction}

If a black hole is perturbed, the space-time geometry will undergo damped oscillations. Such oscillations corresponds to quasinomal modes and the frequencies of such modes are complex. Studies of Quasinormal modes (QNM) of perturbations by gravitational and matter field have taken an important place in black hole physics. QNM's gives information on the stability properties of black holes. Since QNM frequencies depend on the black hole properties such as the mass, angular momentum and charge, they allow a direct way of identifying the space-time parameters. If the radiation due to QNM modes are  detected in the future by gravitational wave detectors, it would be a clear way of identifying the possible charges of black holes. There are extensive studies of QNM's in various blackhole backgrounds in the literature. See the review by Kokkotas et. al. \cite{kok1} for more information.

Due to the the conjecture relating anti-de Sitter(AdS) and conformal field theory (CFT) \cite{aha}, QNM's in AdS spaces have been subjected to intensive investigation. It is conjectured that  the imaginary part of the QNM's which gives the time scale to decay the black hole perturbations corresponds to the time scale of the CFT on the boundary to reach thermal equilibrium. There are  many work on AdS black holes in four and higher dimensions on this subject \cite{chan}\cite{horo} \cite{car1} \cite{moss} \cite{wang}  \cite{kok2}. 

To the authors knowledge, all the work on QNM's of black holes in four and higher dimensions  are numerical except for the massless topological black hole calculation done by Aros et. al. \cite{aros}. However, in 2+1 dimensions, QNM's can be computed exactly due to the nature of the wave equations. In particular,  the well known BTZ black hole \cite{banados} has been studied with exact results \cite{bir1} \cite{bir2} \cite{car2} \cite{abd}. In this paper we take a step further by studying QNM's of a charged black hole in 2+1 dimensions which leads to exact results. To authors knowledge, all the charged blackholes studied for QNM's have been done numerically.

Extensions of the $BTZ$ black hole with charge have lead to many interesting work. The first investigation was done by Banados et. al.\cite{banados}. Due to the logarithmic nature of the electromagnetic  potential, these solutions give rise to unphysical properties\cite{ chan1}. The horizonless static solution with magnetic charge were studied by Hirshmann et.al.\cite{hirsh} and the persistence of these unphysical properties was highlighted by Chan \cite{chan1}. Kamata et.al.\cite{kamata} presented a rotating charged black hole with self (anti-self) duality imposed on the electromagnetic fields. The resulting solutions were asymptotic to an extreme BTZ black hole solution but had diverging mass and angular momentum \cite{chan1}. Clement \cite{clem}, Fernando and Mansouri\cite{fer} introduced a Chern-Simons term as a regulator to screen the electromagnetic potential and obtained horizonless charged particle-like solutions.

In this paper we consider an interesting class of black hole solutions obtained by Chan and Mann \cite{chan2}. The solutions represents static charged black holes with a dilaton field. Furthermore, it has finite mass unlike some of the charged black holes described above. The action considered by Chan and Mann \cite{chan2} is given as follows:
\begin{equation}
S = \int d^3x \sqrt{-g} \left[ R - \frac{B}{2} (\bigtriangledown \phi )^2 -
e^{-4 a \phi} F_{\mu \nu} F^{\mu \nu} + 2 e^{b \phi} \Lambda \right]
\end{equation}
The parameter $ \Lambda$ is  the cosmological constant. ( $\Lambda > 0$ anti-de Sitter and  $\Lambda < 0$ de Sitter).The constants $a$, $b$ and $B$ are arbitrary couplings. $\phi$ is the dilaton field, $R$ is the scalar curvature and $F_{\mu \nu}$ is the Maxwell's field strength. This action is conformally related to the low-energy string action in 2+1 dimensions for  $ B=8, b = 4$ and $a=1$. The black hole in \cite{chan2} could be compared to charged-dilaton black hole in 3+1 dimensions constructed by Gibbons et. al. \cite{gibb} and Grfinkle et. al. \cite{garf}. These black holes have very interesting properties. Furthermore its relation to low-energy string theory makes it an ideal model to study issues in string theory in a simpler setting. 

The paper is presented as follows: In section 2 the black hole solutions are introduced. In section 3 the scalar perturbations are given. In section 4 quasinormal modes for  uncharged black holes are given. In section 5, quasinormal modes for  charged black holes are computed. Finally, the conclusion is given in section 6.

%%%%%%%%%%%%%%%%%%

\section{Static Charged Dilaton Black Hole Solutions in 2+1 Dimensions}

In this section we will give a an introduction to the static charged black hole solutions for the action in eq.(1) obtained by Chan and Mann \cite{chan1}. A family of static solutions with rotational symmetry for the above action were derived in the form,
$$
ds^2 = - \left(-2 M r ^{ 2/N -1} + \frac{ 8 \Lambda r^2}{ (3N-2)N} + 
\frac{ 8 Q^2}{ (2 -N)N} \right)   dt^2$$ 
\begin{equation}
+  \frac{ 4 r^{4/N-2} dr^2}{N^2 \gamma^{4/N} \left((-2 M r ^{ 2/N -1} + \frac{ 8 \Lambda r^2}{ (3N-2)N}  + 
\frac{ 8 Q^2}{ (2 -N)N} \right) }
 + r^2  d \theta^2
\end{equation}
where,
\begin{equation}
k = \pm \sqrt{ \frac{N( 2 -N)}{ 2B}}; \hspace{1.0cm} 4ak = bk = N-2; \hspace{1.0cm} 4a = b;
\end{equation}
The corresponding dilaton field is given by
\begin{equation}
\phi = \frac{2 k}{N} ln \left( \frac{r}{\eta} \right)
\end{equation}
In this paper we will focus on the special class of black hole  with  $N=1, k= -1/4, b=4a=4$. Here the space-time is simplified to,
$$
ds^2= - \left( -2Mr + 8 \Lambda r^2 + 8 Q^2 \right) dt^2 +  \frac{4 r^2 dr^2}{\left(-2Mr + 8 \Lambda r^2 + 8 Q^2 \right)  } + r^2 d \theta^2
$$
\begin{equation}
\phi = k  ln (\frac{r}{\eta}) ; \hspace{1.0cm}F_{rt} = \frac{Q}{r^2}
\end{equation}
The above space-time represents a black hole with two horizons if $M \geq 8 Q \sqrt{\Lambda}$. The two horizons are given by,
\begin{equation}
r_e =  \frac{M + \sqrt{ M^2 - 64 Q^2 \Lambda}}{8 \Lambda}; \hspace{1.0cm}
r_c = \frac{M - \sqrt{ M^2 - 64 Q^2 \Lambda}}{8 \Lambda}
\end{equation}
Here $r_e$ corresponds to the even horizon and $r_c$ corresponds to the Cauchy horizon. The two horizons merge when $ M = 8 Q \sqrt{\Lambda}$ leading to extreme black holes. There  is a time-like singularity at $ r =0$. The  Hawking temperature $T_{Hawking}$. It is given by,
\begin{equation}
T_{Hawking}= \frac{1}{4 \pi} |\frac{dg_{tt}}{dr}| \sqrt{-g^{tt} g^{rr}} |_{r=r_e} = \frac{M}{4 \pi r_e} \sqrt{ 1 - \frac{64 Q^2 \Lambda}{M^2}}
\end{equation}
The temperature  is zero  for the extreme black hole with $M= 8 Q \sqrt{\Lambda}$.

For $b = 4$ and $ B = 8$, the action in eq.(1) has an uncharged-dilaton solution as follows:
$$
ds^2_{uncharged-dilaton \hspace{ 0.2 cm} BH} = - ( 8 \Lambda \delta r - 2 m \sqrt{r})  dt^2 +  \frac{dr^2}{( 8 \Lambda \delta r - 2 m \sqrt{r})  } + \gamma^2 r d \theta^2
$$
\begin{equation}
\phi = -\frac{1}{4} ln ( \frac{r}{\delta})
\end{equation}
Here $\gamma$ is an integrating constant and $m^2$ is the mass per unit length. Interestingly, the above black hole has relation to a $( 1 + 1)$  black hole in string theory found by Mandal et.al. \cite{mandal} (called MSW black hole) given by,
\begin{equation}
ds^2_{MSW} = - ( 1  - \frac{M}{r})  dt^2 +  \frac{ \rho dr^2}{ 4r ( r -M)  } 
\end{equation}
\begin{equation}
\phi = -\frac{1}{2} ln r - \frac{1}{4} ln \rho
\end{equation}
Here, $M$ is the mass and $\rho$ is a constant. The relation between the black hole in eq.(8) and one in eq.(9) goes as follows: When one takes the product  of MSW black hole with $\bf{R}$, one obtain the uncharged-dilaton black hole in eq.(8). MSW black hole has attracted attention in the research community. Exact conformal field theory corresponding to the MSW black hole was found by Witten in \cite{witten}.

As mentioned in the introduction, the Einstein-Maxwell-dilaton action in eq.(1) is dual to the low energy string action in 2 +1 dimensions by the   conformal transformation,
\begin{equation}
g^{String} =  e^{4 \phi} g^{Einstein-Maxwell-dilaton}
\end{equation}
Since in  string theory, different space-time geometries can be related to each other by duality transformations, it is possible to generate charged solutions from uncharged solutions. An extensive review on this subject is given by Horowitz in \cite{horo1}.

It is possible to apply the following  duality transformation to a given  metric in string frame to obtain a charged version.
$$ g_{tt}^{charged} = \frac{ g_{tt}}{[ 1 + (1 + g_{tt}) Sinh^2 \epsilon ]^2}$$
$$A_t^{charged} = - \frac{ ( 1 + g_{tt}) Sinh 2\epsilon )}{2 \sqrt{2} [ 1 + (1 + g_{tt}) Sinh^2 \epsilon ]}$$
\begin{equation}
e^{- 2 \phi(charged)} = e^{ -2 \phi} [ 1 + (1 + g_{tt}) Sinh^2 \epsilon ]
\end{equation}
Here, $g_{tt}$ is in the string frame (or for the uncharged metric). Here $\epsilon$ is an arbitrary parameter. Now, one can take the metric in eq.(8) and perform the duality transformation in eq.(12) to obtain the following dual metric as,
$$
ds^2_{Einstein} = - ( 8 \Lambda \delta r - 2 m \sqrt{r})  dt^2 +  \frac{ \Gamma(r)dr^2 }{( 8 \Lambda \delta r - 2 m \sqrt{r})  } + \gamma^2 r  \Gamma(r)d \theta^2$$
\begin{equation}
e^{-2\phi} = \sqrt{ \frac{r}{\delta} } \Gamma(r)
\end{equation}
Here,
\begin{equation}
\Gamma(r) =  1 + (1 - 8 \Lambda \delta^2) Sinh^2 \epsilon  + \frac{ 2 m \delta Sinh^2 \epsilon}{\sqrt{r}}
\end{equation}
By performing a coordinate transformation given by $\hat{r} = \Gamma(r)^2 r$ and replacing $\hat{r}$ by $r$, yields the charged-dilaton black hole in eq.(5). Hence, the uncharged-dilaton black hole is dual to the static charged dilaton black hole studied in this paper. The duality transformation given in eq.(12) is a part of  $O(2,1)$ symmetry group of the low energy string action as discussed by Sen \cite{sen}. In fact, using another transformation, which is a  part of the symmetry group $O(2,1)$ on the BTZ black hole, Horowitz and Welch obtained another charged black hole in 2 +1 dimensions as,

$$
ds^2_{charged} = - \left( 1 -  \frac{M} {r}\right)  dt^2 +  \left( 1 -  \frac{Q^2} {Mr} \right)^{-1}  dx^2 + \left( 1 -  \frac{M} {r} \right) \left( 1 -  \frac{Q^2} {Mr} \right)^{-1} + \frac{ l^2 dr^2 }{4 r^2  } $$
\begin{equation}
e^{-2\phi} =  ln rl, \hspace{1.0cm} \hat{B}_{xt} = \frac{Q}{r}
\end{equation}
Notice that the transformations used to obtain the two charged black holes are different. However, if $Q =0$, the black hole in eq.(15) corresponds to the one in eq.(8).

%%%%%%%%%%%%%%%%%%%%%%%%%%%%%%%%%%%%%%%%%

\section{Scalar Perturbation of Charged Black Holes}
In this section we will present the equations for the scalar field around the black hole. The general equation for a massive scalar field in curved space-time can be written as,
\begin{equation}
\bigtriangledown ^2 \Phi - \mu ^2 \Phi =0
\end{equation}
which  can be expanded as,
\begin{equation}
\frac{1}{\sqrt{-g}} \partial_{\mu} ( \sqrt{-g} g^{\mu \nu} \partial_{\nu} \Phi ) - \mu^2 \Phi =0
\end{equation}
Here, $\mu$  is the mass of the scalar field. Using the anzatz,
\begin{equation}
\Phi =  e^{- i \omega t} e^{i m \theta} \Psi(r) 
\end{equation}
The radial equation component of eq.(17) is
\begin{equation}
\frac{d}{dr} \left( \frac{D(r)}{2} \frac{d\Psi(r)}{dr} \right) + 2r^2 \left( \frac{\omega^2}{D(r)}   - \frac{m^2}{r^2} - \mu^2 \right)  \Psi(r) =0
\end{equation} 
with the change of variable,
\begin{equation}
 z = \left( \frac{ r - r_e}{ r - r_c} \right)
\end{equation}
the eq.(19) becomes,
\begin{equation}
z(1-z) \frac{d^2 \Psi}{dz^2} + (1-z) \frac{d \Psi}{dz} + P(z) \Psi =0
\end{equation}
Here,
\begin{equation}
P(z) = \frac{A_1}{z} + \frac{A_2}{-1+z} + A_3
\end{equation}
where,
$$A_1=  \frac{\omega^2 r_e^2}{ 16 (r_e - r_c)^2 \Lambda^2}$$
$$
A_2 = \frac{ 8m^2 \Lambda - \omega^2} { 16 \Lambda^2}
$$
\begin{equation}
A_3 = - \frac{r_c^2 \omega^2}{16 (r_e- r_c)^2 \Lambda^2}
\end{equation}
Note that in the new coordinate system, $z=0$ corresponds to the horizon and $z =1$ corresponds to the infinity.

%%%%%%%%%%%%%%%%%%
\section{Quasinormal Modes of an Uncharged Dilaton Black Hole}

When a black hole is perturbed, the response will  undergo three stages. The first stage will depend on the initial condition. The second stage corresponds to oscillations due to complex frequencies. These modes are called quasinormal modes. The last stage corresponds to late time decay which goes as a power law.

In this paper, our main goal is to compute the frequencies of the quasi normal modes of the scalar perturbations.  In this work, the quasinormal modes corresponds to the wave solutions of eq.(17) with appropriate boundary conditions imposed. There are two boundary conditions. At the horizons, it is imposed that the wave is purely ingoing.  The other boundary condition is at the asymptotic infinity. For asymptotically flat space-time, the boundary condition is for the wave to be purely out going. For asymptotically AdS scape-times, various  boundary conditions are chosen. One of the chosen condition is the Dirichlet boundary condition and the other is to make the energy momentum flux to be zero at infinity\cite{bir2}. In this paper, we will choose  the Dirichlet condition.

First,  the uncharged black string solution with $Q=0$ is considered. This  solution has a horizon at  $r_h= M/4 \Lambda$.  To compute the modes we will choose scalar perturbation with $\mu =0$. For this case, $P(z)$  in eq.(22) is,
$$P(z) = \frac{A_1}{z} + \frac{A_2}{-1+z}$$
$$A_1 =  \frac{\omega^2}{ 16 \Lambda^2}$$
\begin{equation}
A_2 = \frac{ 8m^2 \Lambda - \omega^2} { 16 \Lambda^2}
\end{equation}
Now, with the  definition of  $\Psi$ as,
\begin{equation}
\Psi(z) = z^{\alpha} (1-z)^{\beta} F(z)
\end{equation}
the radial equation given in eq.(21) becomes,
\begin{equation}
z(1-z) \frac{d^2 F}{dz^2} + \left(1 + 2 \alpha - (1+ 2 \alpha + 2\beta )z \right) \frac{d F}{dz} + \left(\frac{\bar{A_1}}{z} + \frac{\bar{A_2}}{-1+z} + \bar{A_3}     \right) F =0
\end{equation}
where,
$$\bar{A_1} = A_1 + \alpha^2$$
$$ \bar{A_2} = A_2 + \beta - \beta^2$$
\begin{equation}
\bar{A_3} = -(\alpha+ \beta)^2
\end{equation}
Equation(27)  resembles the hypergeometric differential equation, \cite{math},
\begin{equation}
z(1-z) \frac{d^2 F}{dz^2} + (c  - (1+a + b )z) \frac{d F}{dz} -ab  F =0
\end{equation}
By comparing the coefficients of eq.(26) and eq.(28),  the following identities are obtained.
\begin{equation}
c = 1+ 2 \alpha
\end{equation}
\begin{equation}
a+b = 2 \alpha + 2 \beta
\end{equation}
\begin{equation}
\bar{A_1}= A_1 + \alpha^2 =0; \Rightarrow \alpha= \frac{ i \omega}{ 4 \Lambda}
\end{equation}
\begin{equation}
\bar{A_2} = A_2 + \beta - \beta^2=0
\end{equation}
\begin{equation}
 ab = -\bar{A_3} = (\alpha + \beta)^2  
\end{equation}
From eq.(30) and eq.(33),
\begin{equation}
a= b= \alpha + \beta
\end{equation}
The hypergeometric function $F(z)$  has a solution  given by \cite{math},
\begin{equation}
F(a,b;c;z) = \frac{\Gamma(c)} {\Gamma(a) \Gamma(b)} \Sigma \frac{ \Gamma(a+n) \Gamma( b+n)}{ \Gamma(c+n)}  \frac{z^n}{n!}
\end{equation}
The  radius of convergence of $F$  is the  unit circle $|z| =1$.There are number of linear transformations for F(a,b;c;z) as given in \cite{math} which helps to write the solution in the form,
$$
\Psi(z) = z^{\alpha} (1-z)^{\beta} \frac{ \Gamma(c) \Gamma(c-a-b)}{\Gamma(c-a) \Gamma(c-b)} F(a,b;a+b-c+1;1-z) $$
\begin{equation}
+ z^{\alpha} (1-z)^{\beta} (1-z)^{c-a-b}\frac{ \Gamma(c) \Gamma(a+b-c)}{\Gamma(a) \Gamma(b)} F(c-a,c-b;c-a-b+1;1-z)
\end{equation}
To impose the vanishing of the wave at $r \rightarrow \infty$, (or at $z=1$), both terms in the above expansion should be zero. At $z=1$ the second term vanish. The first term vanish only at the poles of $\Gamma(c-a)$ and $\Gamma(c-b)$. Note that the  Gamma function $\Gamma(x)$ has poles at $x=-n$ for $ n=0,1,2..$. Hence for the  function $\Psi(z)$ to vanish at $z=1$ the following additional restriction has to be applied.
\begin{equation}
c-a =  -n
\end{equation}
\begin{equation}
c -b = -n; \hspace{1.0cm} n=0,1,2,....
\end{equation}
By combining eq.(29), eq.(31), eq.(34) and eq.(37) leads to,
\begin{equation}
1+n= - \frac{ i \omega}{ 4 \Lambda} + \beta
\end{equation}
By Combining eq.(24) and eq.(32) leads to,
\begin{equation}
\frac{m^2}{  2 \Lambda} - \frac{ \omega^2}{ 16 \Lambda^2} = - \beta + \beta^2
\end{equation}
eliminating $\beta$ from the above equations(39) and (40) leads to the quasinormal frequency $\omega$ as,
\begin{equation}
\omega =  \frac{2 i}{ 2n +1} \left( 2 \Lambda n (1+n) - m^2 \right)
\end{equation}

Hence,  the QNM's of this black holes are  pure-imaginary. In a recent article, Berti et. al. \cite{kok2} showed that the Schwarzchild-AdS black holes to have  pure imaginary QNM's for gravitational and electromagnetic perturbations. Here,  the frequency behaves as the inverse of the AdS radius for small values of $m$. It is noted that the Hawking temperature of the uncharged black hole, $T_H =\Lambda/\pi$  is independent of the black hole mass and so do  $|\omega|$. $|\omega|$ $\sim \Lambda$ for small $m$ values and therefore there is a  linear relation between $\omega$ and $T_H$. 

%%%%%%%%%%%%%%%%%%%%%%%%%%%
\section{Quasinormal Modes of Charged Black Holes}

In this section we will focus on the charged black hole with non-zero charge $Q$. The function $P(z)$ for this case is,
$$P(z) = \frac{A_1}{z} + \frac{A_2}{-1+z}+ A_3$$
$$A_1 = \frac{r_e^2 \omega^2}{ 16 (r_e - r_c)^2\Lambda^2}$$
$$
A_2 = \frac{ 8m^2 \Lambda - \omega^2} { 16 \Lambda^2}$$
\begin{equation}
A_3= - \frac{ r_c^2 \omega^2}{ 16 (r_e - r_c)^2 \Lambda^2}
\end{equation}
Now, with the  definition of  $\Psi$ as,
\begin{equation}
\Psi(z) = z^{\alpha} (1-z)^{\beta} F(z)
\end{equation}
the radial equation given in eq.(21) becomes,
\begin{equation}
z(1-z) \frac{d^2 F}{dz^2} + \left(1 + 2 \alpha - (1+ 2 \alpha + 2\beta )z \right) \frac{d F}{dz} + \left(\frac{\bar{A}}{z} + \frac{\bar{B}}{-1+z} + \bar{C}\right) F =0
\end{equation}
where,
$$\bar{A_1} = A_1 + \alpha^2$$
$$ \bar{A_2} = A_2 + \beta - \beta^2$$
\begin{equation}
\bar{A_3} = A_3 - (\alpha+ \beta)^2
\end{equation}
As was for the uncharged case,  the above equation resembles the hypergeometric differential equation of the format given in eq.(26) in the previous section. By comparing the coefficients of eq.(26) and eq. (44), one can obtain,
\begin{equation}
c = 1+ 2 \alpha
\end{equation}
\begin{equation}
a+b = 2 \alpha + 2 \beta
\end{equation}
\begin{equation}
\bar{A_1}=A_1 + \alpha^2 =0; \Rightarrow \alpha= \frac{\pm  i \omega r_e }{ 4 \Lambda ( r_e - r_c) }
\end{equation}
\begin{equation}
\bar{A_2} = A_2 - \beta + \beta^2=0
\end{equation}
\begin{equation}
 ab = -\bar{A_3} = (\alpha + \beta)^2 - A_3 
\end{equation}
From eq.(47) and eq.(50),
$$a= \alpha + \beta + i \sqrt{|A_3|}$$
\begin{equation}
b= \alpha + \beta - i \sqrt{|A_3|}
\end{equation}
Similar to the uncharged case, the boundary conditions on the solution $\Psi(z)$ will impose the restriction on $a,b,c$ as,
$$c-a =  -n$$
\begin{equation}
c -b = -n; \hspace{1.0cm} n=0,1,2,3....
\end{equation}
By combining eq.(46), eq.(51) and eq.(52) leads to,
\begin{equation}
 1+n= \pm \alpha + \beta \pm i \sqrt{|A_3|}
\end{equation}
Due to the possibility of $``\pm''$ in the eq.(52), there are four possibilities for $\kappa$  given in the following equations;
\begin{equation}
 \pm \alpha \pm i \sqrt{|A_3|} =  \kappa \omega
\end{equation}
where,
\begin{equation}
\kappa = \pm 1, \pm \left( \frac{r_e + r_c}{r_e - r_c} \right)
\end{equation} 
Hence eq.(53) can be rewritten as,
\begin{equation}
1+n = \beta + i \frac{\kappa \omega}{ 4 \Lambda}
\end{equation} 
By eliminating $\beta$ from eq.(49) and eq.(56) one can obtain  the quasinormal frequency $\omega$ as,
\begin{equation}
\omega= \pm i \frac{2}{ ( 1- \kappa^2)} \left(  -\kappa \Lambda(1+2 n) + \sqrt{\Lambda} \sqrt{ 2m^2 	(\kappa^2 -1) + \Lambda(\kappa^2 + 4 n  (1+n))}    \right); \hspace{0.5cm} \kappa =   \pm \left( \frac{r_e + r_c}{r_e- r_c}  \right)
\end{equation}
The $\omega$ for $\kappa = \pm 1$ is the same  as for the uncharged case. Therefore  it  was eliminated without loss of generality.  $\omega$ will always be pure imaginary since $\kappa >1$.

\section{Conclusion}
We have computed the exact values of the Quasinormal modes of the static dilaton black holes in 2+1 dimensions. For both charged and uncharged black holes, the QNM's are pure imaginary which is a novel property in contrast to the QNM's of the BTZ black hole in 2+1 dimensions. To the authors knowledge, this is the first exact computation of QNM's of a charged black hole.

The work here motivate studies in various directions. The first is to study the AdS/CFT relation in terms of QNM's. It is conjectured that  the imaginary part of the QNM's which gives the time scale to decay the black hole perturbations corresponds to the time scale of the CFT on the boundary to reach thermal equilibrium. Since the computation on the AdS side is already  done it is worth studying the CFT on the boundary of these black holes. The  uncharged black hole and the charged black hole considered in this paper are dual to each other. It would be interesting to study if there is a relation between them which highlights the duality property.

As was noticed for the uncharged black hole, the quasinormal frequency has a linear relation with the Hawking temperature $T_H$ and is independent of the mass of the black hole. I like to note that Horowitz et.al.\cite{horo} did a numerical computation to show that the imaginary part of the quasinormal frequency ($\omega_I$) scale with the horizon radius $r_+$ as $ \omega_I \sim \frac{r_+}{\eta}$ where $\eta$ is called the Choptunik parameter \cite{chop}. It is worthwhile to study this relation in terms of the dilaton black hole given in this paper. Note that such a relation was established for the BTZ black hole in \cite{bir1}.

The static charged black hole considered here has similar properties to 3+1 dimensional Reissner-Nordstrom black hole in 3 +1 dimensions which is well studied. I believe the dilaton  solutions considered is an  excellent model to study the properties of charged black holes in a simpler setting.

After publishing the paper, we found out that there are three papers which are related to this work which done by Mann and Chan. These works are referenced in \cite{mann1} \cite{mann2}\cite{mann3}.


\begin{thebibliography}{99}
%\bibitem
\bibitem{kok1} K.D. Kokkotas, B.G. Schmidt, Living Rev. Relativ. {\bf2} (1999) 2
\bibitem{aha} O. Aharony, S.S. Gubser, J. Maldacena, H. Ooguri \& Y. Oz, ``Large N Field Theories'', hep-th/9905111.
\bibitem{chan} J. Chan \& R. Mann, Phys. Rev. {\bf D55} (1997) 7546; Phys. Rev. {\bf 59} (1999) 064025
\bibitem{horo} G.T. Horowitz \& V. E. Hubeny, Phys. Rev. {\bf D62} (2000) 024027.
\bibitem{car1} V. Cardoso \& J.P.S. Lemos, Phys. Rev. {\bf D64} (2001) 084017
\bibitem{moss} I.G. Moss \& J.P. Norman, Class. Quan. Grav. {\bf 19} (2002), 2323
\bibitem{wang} B. Wang, C.Y.Lin \& E. Abdalla, Phys. Lett {\bf B481} (2000) 79
\bibitem{kok2} E. Berti \& K.D. Kokkotas, Phys. Rev {\bf D67} (2003) 064020
\bibitem{aros} R. Aros, C. Martinez and R. Troncoso \& J. Zanelli, Phys. Rev. {\bf D67} (2002) 044014
\bibitem{bir1} D. Birmingham, Phys. Rev {\bf D64} (2001) 064024
\bibitem{bir2} D. Birmingham, I. Sachs \& S.N. Solodukhin, Phys. Rev .Lett. 88 (202) 151301.
\bibitem{car2} V. Cardoso \& J.P.S. Lemos, Phys. Rev. {\bf D63} (2001) 124015
\bibitem{abd} E. Abdalla, B. Wang, A. Lima-Santos \& W.G. Qiu, Phys. Lett. {\bf B38} (2002) 435
\bibitem{banados} M. Ba\~{n}ados, C. Teitelboim, J. Zanelli,
 Phys. Rev. Lett. {\bf 69} (1992) 1849; M. Ba\~{n}ados, M.
Henneaux, C. Teitelboim, J.
Zanelli, Phys. Rev. D {\bf 48} (1993) 1506.
\bibitem{chan1} K.C.K. Chan, Phys. Lett. {\bf B373} (1996) 296
\bibitem{hirsh} E. W. Hirshmann, D.L. Welch, Phys. Rev. {\bf D53} (1996) 5579.

\bibitem{kamata}, T. Koikawa, Phys. Lett{ \bf B 353} (1995) 196
\bibitem{clem} G. Clement, Phys. Lett. {\bf B 367} (1996) 70.

\bibitem{fer} S. Fernando, F. Mansouri, Commun. Math. And Theo. Phys. 1 (1998) 14.

\bibitem{chan2}K.C.K. Chan, R.B. Mann ,  Phys. Rev.  {\bf D50} (1994) 6385.
\bibitem{gibb} G. W. Gibbons, K. Madea,  Nucl. Phys. {\bf B298}   (1988) 741
\bibitem{garf} D. Garfinkle, G. T. Horowitz ,  A. Strominger,    Phys. Rev  {\bf D43}     (1991) 3140
\bibitem{mandal} G. Mandal, A. M. Senguptha and S. R. Wadia, Mod. Phys. Lett. {\bf A6}, (1991) 1685
\bibitem{witten} E. Witten, Phys. Rev. {\bf D44}, (1991) 314 



\bibitem{horo1} G. Horowitz, { \it ``The Dark Side of String Theory: Black Holes and Black Strings''}, hep-th/9210119.
\bibitem{sen} S. Hassan and A. Sen, Nucl. Phys. {\bf B 375}, (1992) 103.

\bibitem{horo2} G. Horowitz and D. Welch, Phys.Rev.Lett. {\bf 71} (1993) 328-331

\bibitem{math} ``Handbook of Mathematical Functions'', M. Abramowitz and A. Stegun, Dover, (1977)

\bibitem{chop} M.W. Choptunik, Phys. Rev. Lett 70 (1993) 9

\bibitem{mann1} S. F. J. Chan and R. B. Mann, Phys Rev. {\bf D 55} (1997) 7546-7562

\bibitem{mann2} S. F. J. Chan and R. B. Mann, Phys Rev. {\bf D 59 } (1999) 064025

\bibitem{mann3} B. Wang, E. Abdalla  and R. B. Mann, Phys Rev. {\bf D 65} (2002) 084006





\end{thebibliography}
\end{document}